\documentclass[10pt,letterpaper]{article}
\usepackage[top=0.85in,left=2.75in,footskip=0.75in,marginparwidth=2in]{geometry}

\usepackage[utf8]{inputenc}

\usepackage{cite}

\usepackage{nameref,hyperref}

\usepackage[right]{lineno}

\usepackage{microtype}
\DisableLigatures[f]{encoding = *, family = * }

\raggedright
\usepackage{changepage}

\usepackage[aboveskip=1pt,labelfont=bf,labelsep=period,singlelinecheck=off]{caption}

\makeatletter
\renewcommand{\@biblabel}[1]{\quad#1.}
\makeatother

\usepackage{lastpage,fancyhdr,graphicx}
\usepackage{epstopdf}
\usepackage{color}

\definecolor{Gray}{gray}{.25}

\usepackage{graphicx}

\usepackage{sidecap}

\usepackage{wrapfig}
\usepackage[pscoord]{eso-pic}
\usepackage[fulladjust]{marginnote}
\reversemarginpar

\begin{document}
\vspace*{0.35in}

\begin{adjustwidth}{-1.5in}{0in}
\begin{flushleft}
{\Large
\textbf{\newline{CSL Collapse Model Mapped with the Spontaneous Radiation}}
}
\newline
\\
K. Piscicchia\textsuperscript{1,2,*},
A. Bassi\textsuperscript{3,4},
C. Curceanu\textsuperscript{2,1},
R. Del Grande\textsuperscript{1},
S. Donadi\textsuperscript{5},
B.C. Hiesmayr\textsuperscript{6},
A. Pichler\textsuperscript{7}
\\
\bigskip
$^{1}$ CENTRO FERMI - Museo Storico della Fisica e Centro Studi e Ricerche "Enrico Fermi", 00184 Rome, Italy;
\\
$^{2}$ Istituto Nazionale di Fisica Nucleare (INFN), Laboratori Nazionali di Frascati, 00044 Frascati, Italy;
\\
$^{3}$ Department of Physics, University of Trieste, 34151 Miramare-Trieste, Italy;\\
$^{4}$ Istituto Nazionale di Fisica Nucleare, Sezione di Trieste, Via Valerio 2, 34127 Trieste, Italy;\\
$^{5}$ Institute of Theoretical Physics, Ulm University, Albert-Einstein-Allee 11 D, 89069 Ulm, Germany;\\
$^{6}$ Faculty of Physics, University of Vienna, Boltzmanngasse 5, 1090 Vienna, Austria;\\
$^{7}$ Stefan-Meyer-Institut f\"ur Subatomare Physik, 1090 Vienna, Austria;
\bigskip
* kristian.piscicchia@lnf.infn.it

\end{flushleft}

\section*{Abstract}
In this paper, new upper limits on the parameters of the Continuous Spontaneous Localization (CSL) collapse model are extracted. To this end, the X-ray emission data collected by the IGEX collaboration are analyzed and compared with the spectrum of the spontaneous photon emission process predicted by collapse models.
 This study allows the obtainment of the most stringent limits within a relevant range of the CSL model parameters, with respect to any other method. The collapse rate $\lambda$ and the correlation length $r_C$ are mapped, thus allowing the exclusion of a broad range of the parameter space.

\end{adjustwidth}
\begin{adjustwidth}{-1.0in}{0in}
\section{The CSL Collapse Model}

Collapse models are phenomenological models introduced to solve the measurement problem of quantum mechanics and explain the quantum-to-classical transition \cite{review}. According to these models, the linear and unitary evolution given by the Schr\"odinger equation is modified by adding a non-linear term and the interaction with a stochastic noise field. These modifications have two very important consequences: (i) they lead to the collapse of the wave function of the system in space (localization mechanism) and (ii) the collapse effects get amplified with the mass of the system (amplification mechanism). The combination of these two properties guarantees that macroscopic objects always have well defined positions, explaining why we do not observe quantum behaviour at the macroscopic level. On the other hand, for microscopic systems, the effect of the non-linear interaction with the noise field is very small and their dynamics is dominated by the Schr\"odinger evolution. 
Due to the presence of the non-linear interaction with the noise field, collapse models predict slight deviations from the standard quantum mechanics predictions \cite{fu}. 

The analysis discussed in this work sets limits on the characteristic parameters of the Continuous Spontaneous Localization (CSL) model \cite{ghi,pear,pear2}, which is one of the most relevant and well-studied collapse models in the literature.
In the CSL model, the state vector evolution is described by a~modified Schr\"odinger equation which contains, besides the standard Hamiltonian, non-linear and stochastic terms, characterized by the interaction with a continuous set of independent noises $w(\textbf{x},t)$ (one for each point of the space, which is why this set is often referred to as ``noise field'') having zero average and white correlation in time, i.e., $E[w(\textbf{x},t)]=0$ and $E[w(\textbf{x},t)w(\textbf{y},s)]=\delta(\textbf{x}-\textbf{y})\delta(t-s)$ where $E[...]$ denotes the average over the noises. 
Two phenomenological parameters ($\lambda$ and $r_C$) are introduced in the model. The parameter $\lambda$ has the dimensions of a rate and sets the strength of the collapse, while $r_C$ is a correlation length which determines the spatial resolution of the collapse: for superposition with size much smaller than $r_C$, the collapse is much weaker compared to the case when the superposition has a delocalization much larger than $r_C$. 
The originally proposed values for $\lambda$ and $r_C$ are \cite{ghi} \mbox{$\lambda = 10^{-16}$ s$^{-1}$}, $r_C = 10^{-7}$ m. Higher values for $\lambda$ were however put forward \cite{Adler}, up to $\lambda = 10^{-8\pm2}$ s$^{-1}$.

The interaction with the noise field  causes an extra emission of electromagnetic radiation  for charged particles \cite{fu}, which is not predicted by standard quantum mechanics. Such an effect is known as \emph{spontaneous radiation} emission. We show that the measurement of the radiation allows for a~mapping of the two relevant parameters $\lambda$ and $r_C$ (see also Ref. \cite{curcepisc}) into a two-dimensional parameter space, i.e., we can present an exclusion plot. This gives a considerable reduction of the possible values in the parameter space of collapse models.

\section{The Collapse Rate Parameter \boldmath {$\lambda$}}\label{lambda}

The energy distribution of the spontaneous radiation, emitted as a consequence of the interaction of free electrons with the collapsing stochastic field, was first calculated by Fu \cite{fu} and later on studied in more detail in \cite{adlram,adlbas,dondec}, in the framework of the non-relativistic CSL model. If the stochastic field is assumed to be a white noise, coupled to the particle mass density (mass proportional CSL model), the~spontaneous emission rate is given by:

\begin{equation}\label{furate1}
\frac{d\Gamma (E)}{dE} = \frac{e^2 \lambda}{4\pi^2 r_C^2 m_N^2 E},
\end{equation}
where $e$ is the charge of the proton, $m_N$ represents the nucleon mass and $E$ is the energy of the emitted photon. In the non-mass proportional case, the rate takes the expression:

\begin{equation}\label{furate2}
\frac{d\Gamma (E)}{dE} = \frac{e^2 \lambda}{4\pi^2 r_C^2 m_e^2 E},
\end{equation}
with $m_e$ the electron mass.

Using the measured radiation emitted in an isolated slab of Germanium \cite{miley} corresponding to an energy of 11 keV, and comparing it with the predicted rate in Equations (\ref{furate1}) and (\ref{furate2}), Fu extracted the following upper limits on $\lambda$ for the two cases:

\begin{equation}\label{fulimit}
\lambda \leq 2.20 \cdot  10^{-10} s^{-1} \;\;\;\mbox{mass prop.},
\end{equation}
\unskip
\begin{equation}\label{fulimit1}
\lambda \leq 0.55 \cdot  10^{-16} s^{-1} \;\;\;\mbox{non-mass prop.},
\end{equation}

assuming that the correlation length value is $r_C = 10^{-7}$ m. In his estimate, Fu considered the contribution to the spontaneous X-ray emission of the four valence electrons in the Germanium atoms. Such electrons can be considered as \emph{quasi-free}, since their binding energy (of the order of $\sim$10 eV) is much less than the emitted photons' energy.
In Ref. \cite{Adler}, the author argues that an erroneous value for the fine structure constant is used in Ref. \cite{fu}. This correction is taken into account in the analysis described in Section \ref{new}.  
Further, the preliminary TWIN data set \cite{miley} used by Fu to estimate the upper limit on $\lambda$ turned out to be underestimated by a factor of about 50 at 10 keV. 

A new analysis was performed in Ref. \cite{mullin}. Based on the improved data presented in Ref. \cite{collett}, the limits corresponding to the footnote [7] in Ref. \cite{mullin}, for the cases of mass proportional and non-mass proportional CSL models, were:

\begin{equation}\label{lim22}
\lambda \leq 8 \cdot  10^{-10} s^{-1} \;\;\;\mbox{mass prop.},
\end{equation}
\begin{equation}\label{lim33}
\lambda \leq 2 \cdot  10^{-16} s^{-1} \;\;\;\mbox{non-mass prop.}.
\end{equation}

\section{A New Limit on \boldmath{$\lambda$}}\label{new}

In this work, the X-ray emission spectrum measured by the IGEX experiment \cite{igex1} is analysed in order to set a more stringent limit on the collapse rate parameter $\lambda$. IGEX is a low-background experiment based on low-activity Germanium detectors, originally dedicated to the neutrinoless double beta decay ($\beta \beta 0 \nu$) research. The published data set \cite{igex2} refers to 80 kg day exposure, and was conceived to search for a dark matter WIMPs signal that originated from elastic scattering, producing Ge nuclear recoil
.

For the measurement in Ref. \cite{igex2}, one of the IGEX detectors of 2.2 kg (active mass of about 2 kg) was used. The detector, the cryostat and the shielding were fabricated following ultra-low background techniques, in order to minimize the radionuclides emission, which represents the main background source in the measured X-ray spectrum (shown in Figure \ref{fit} as a black distribution). Moreover, a~cosmic muon veto covered the top and the sides of the shield. The experiment had an~overburden of \mbox{2450 m.w.e.
,} reducing the muon flux to the value of $2 \cdot 10^{-7}$ cm$^{-2}$ s$^{-1}$. The two main sources of inefficiency are represented by the muon veto anti-coincidence and the pulse shape analysis. The probability of rejecting non-coincident events with the muon veto was found to be less than 0.01. The loss of efficiency introduced by the pulse shape analysis resulted to be negligible for events above 4 keV.   
\begin{figure}[!h]
\begin{adjustwidth}{-1.0in}{0in}
\centering
\includegraphics[width=9cm]{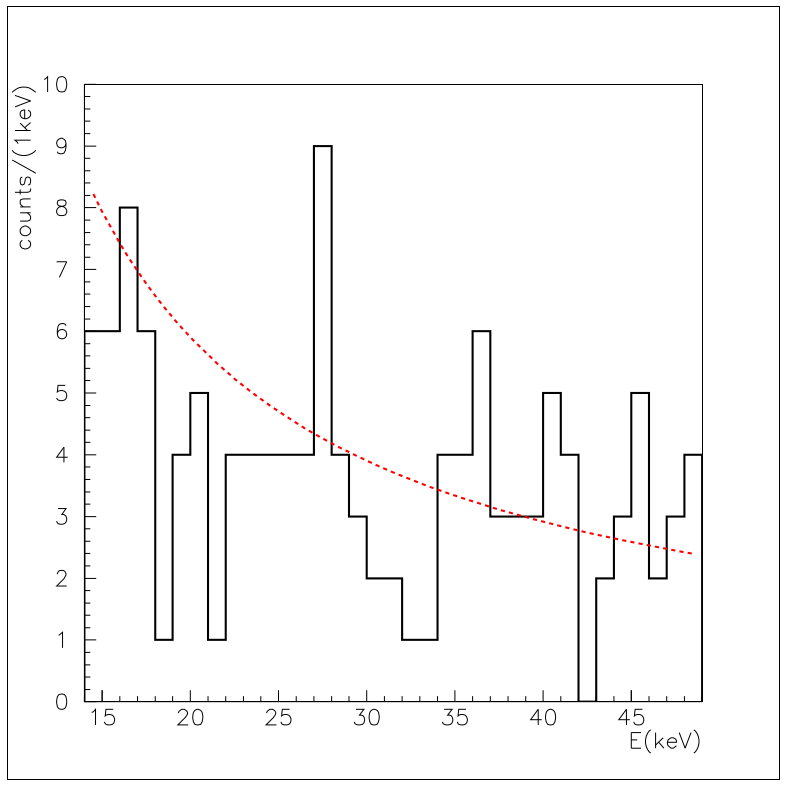}
\caption{
Fit of the X-ray emission spectrum measured by the IGEX experiment \cite{igex1,igex2}, using the theoretical fit function Equation (\ref{fitfunc}). 
The black line corresponds to the experimental distribution;
the red dashed line represents the fit. See the text for more details.}
\label{fit}
\end{adjustwidth}
\end{figure}

The X-ray spectrum (Figure \ref{fit}) ranges in the interval ($4.5\div 48.5$) keV, which is compatible with the non-relativistic assumption for electrons, used to derive Equations (\ref{furate1}) and (\ref{furate2}).

\subsection{The Data Analysis: Procedure and Results}\label{fitpar}

The X-ray experimental spectrum published in \cite{igex2} is compared with the predicted rate Equations (\ref{furate1}) and (\ref{furate2}), by taking into account the spontaneous emission of the 30 outermost electrons of the Ge atoms considered as \emph{quasi-free}. We restricted our analysis to the energy range $\Delta E =$ ($14.5\div 48.5$) keV of the experimental spectrum \cite{igex2}, for which the binding energy of the lower lying electronic orbit (the 2s orbit) is still one order of magnitude lower than 14.5 keV, justifying the \emph{quasi-free} hypothesis.

The X-ray spectrum is fitted in the interval $\Delta E$ by minimising a $\chi^2$ function. The expected number of counts for each bin of 1 keV is assumed to be described by the theoretical prediction Equations (\ref{furate1}) and (\ref{furate2}):

\begin{equation}\label{fitfunc}
\frac{d\Gamma (E)}{dE} = \frac{\alpha(\lambda)}{E}.
\end{equation}

The $\chi^2$ minimisation presumes that the bin contents $y_i$ (number of counts in the energy bin $E_i$) follow Gaussian distributions. Strictly speaking, the $y_i$s are Poissonian stochastic variables; nevertheless, the approximation is reasonable for $y_i \ge 5$; this constraint is then used for the fit. The result of the fit is shown in Figure \ref{fit} (red dashed line). For the free parameter of the fit, the minimization gives the value $\alpha(\lambda) = 115 \pm 17$, corresponding to a reduced $\chi^2/(n.d.f.-n.p.) = 0.9$. $n.d.f.$ represents the number of degrees of freedom, $n.p.$ is the number of free parameters of the fit. 
$\alpha(\lambda)$ is also considered to follow a~Gaussian distribution with a good approximation. An upper limit can then be set as $\alpha(\lambda) \leq 143$ with a probability of 95\%. Correspondingly, an upper limit on the parameter $\lambda$ can be extracted using Equations (\ref{furate1}) and (\ref{furate2}):    
\begin{equation}\label{fu1}
\frac{d\Gamma (E)}{dE} = c \, \frac{e^2 \lambda}{4\pi^2 r_C^2 m^2 E}  \leq \frac{143}{E},
\end{equation}
where the factor $c$ is given by:

\begin{equation}
c = \left(8.29 \times 10^{24}\, \frac{\textrm{atoms}}{\textrm{kg}}\right)\cdot(80 \, \textrm{kg \ day})\cdot \left(8.64 \times 10^4 \, \frac{\textrm{n. of seconds}}{\textrm{day}}\right)\cdot(30),
\end{equation}
the first bracket accounts for the particle density of Germanium, the second represents the amount of emitting material expressed in kg day, the third term is the number of seconds in one day and 30 represents the number of spontaneously emitting electrons for each Germanium atom. Applying Equation (\ref{fu1}), the following upper limits for the reduction rate parameter are obtained, with a probability of 95\%:
\begin{equation}\label{lim2}
\lambda \leq 8.1 \cdot  10^{-12} s^{-1} \;\;\;\mbox{mass prop.},
\end{equation}
\begin{equation}\label{lim3}
\lambda \leq 2.4 \cdot  10^{-18} s^{-1} \;\;\;\mbox{non-mass prop.}.
\end{equation}

In order to obtain the limits in Equations (\ref{lim2}) and (\ref{lim3}), two implicit assumptions are made on the experimental input \cite{igex2}. First, the measured spectrum is assumed to be background free, that is to say that the upper limit on $\lambda$ corresponds to the case in which all the measured X-ray emission would be produced by spontaneous emission processes. This ansatz is conservative, and is imposed by our ignorance regarding the contribution from known emission processes to the measured rate. The~second assumption, which is consistent with the analysis presented in Ref. \cite{igex2}, is that the detector efficiency, in the range  $\Delta E$, is one, and that the un-efficiencies which are introduced by the muon veto anticoincidence and the pulse shape analysis, performed to extract the experimental spectrum in Ref.~\cite{igex2}, are very small for events above 4 keV.

Having in mind these assumptions, the measured X-ray counts in the range $\Delta E$ can be re-analysed in terms of their low-events Poissonian statistics. The number of counts $y_i$s in each energy bin $E_i$ can be considered as independent stochastic variables following the distributions:
\begin{equation}\label{eq2}
G(y_i|P, \Lambda_i) = \frac{\Lambda_i^{y_i}e^{-\Lambda_i}}{y_i!},
\end{equation}
where $P$ denotes the Poisson distribution function. The expected numbers of counts per bin $\Lambda_i$ are indicated with capital letters, not to be confused with the spontaneous collapse rate $\lambda$.
Let us define:
\begin{equation}\label{eq3}
y = \sum_{i=1}^n y_i \qquad,\qquad \Lambda = \sum_{i=1}^n \Lambda_i
\end{equation}
where $n$ is the total number of 1 keV bins in the range $\Delta E$, $y$ and $\Lambda$ are the total number of counts and the expected number of total counts, respectively. Here, $y$ is distributed according to a Poissonian of parameter $\Lambda(\lambda)$, where the dependence on the collapse rate parameter, which follows the theoretical input, was explicitly indicated.

According to the Bayes theorem, the probability distribution function of $\Lambda(\lambda)$, given the measured $y$, assuming a uniform prior, is given by:
\begin{equation}\label{eq4}
G'(\Lambda|G(y|P,\Lambda)) \propto \Lambda(\lambda)^{y}e^{-\Lambda(\lambda)},
\end{equation}
which means that $G'(\lambda)$ is proportional to a gamma probability distribution. Due to the assumption that the background is negligible, $\Lambda(\lambda)$ also represents the expected number of total signal counts $y_s$, where $y_s$ is a Poissonian variable. Thus, according to Equation (\ref{fu1}):
\begin{equation}\label{eq5}
\Lambda(\lambda) = y_s +1 = \sum_{i=1}^n c \, \frac{e^2 \lambda}{4\pi^2 r_C^2 m^2 E_i} +1 = \sum_{i=1}^n  \frac{\alpha(\lambda)}{E_i} +1 .
\end{equation} 

Substituting Equation (\ref{eq5}) for Equation (\ref{eq4}), the probability distribution function for the collapse rate parameter can then be obtained:

\begin{equation}\label{eq6}
G'(\lambda|G(y|P,\Lambda)) \propto \left( \sum_{i=1}^n  \frac{\alpha(\lambda)}{E_i} +1 \right)^{y}e^{- \left(\sum_{i=1}^n  \frac{\alpha(\lambda)}{E_i} +1 \right)},
\end{equation}
{where the measured total number of counts is $y = 130$. Calculating the cumulative distribution function:}

\begin{equation}\label{eq7}
\int_0^{\lambda_0} G'(\lambda|G(y|P,\Lambda)) \,  \mathrm{d}\lambda,
\end{equation}
the following upper limits can be obtained on the collapse rate parameter, setting $r_C$ to the value \mbox{$10^{-7}$ m}, corresponding to a probability level of 95\%
\begin{equation}\label{eq8}
\lambda \leq 6.8 \cdot  10^{-12}~s^{-1} \;\;\;\mbox{mass prop.},
\end{equation}
\begin{equation}\label{eq9}
\lambda \leq 2.0 \cdot  10^{-18}~s^{-1} \;\;\;\mbox{non-mass prop.}.
\end{equation}



\section{Mapping CSL Parameters Space}\label{mapping}

In Figure \ref{csl}, we present the mapping of the $\lambda - r_C$ parameters of the CSL model, where the originally proposed theoretical values are shown, together with our results. The region excluded by theoretical arguments is represented in gray. 
This theoretical bound (see Ref. \cite{angelo}) is obtained by requiring that a single-layered graphene disk of radius $\sim$0.01 mm is localized within $\sim$10 ms (these are the minimum resolution and perception time of the human eye, respectively).

The region excluded by this analysis is shown in cyan for the non-mass proportional case and in magenta for the mass proportional case. Figure \ref{csl} can be compared with Figure \ref{csl} in Ref. \cite{matteo}, where the mapping is obtained using other measurements. It is interesting to note that, for a collapse induced by a white noise, the allowed parameter space is confined to a drastically reduced region. 

\begin{figure}[!h]
\begin{adjustwidth}{-1.0in}{0in}

\centering
\includegraphics[width=9cm]{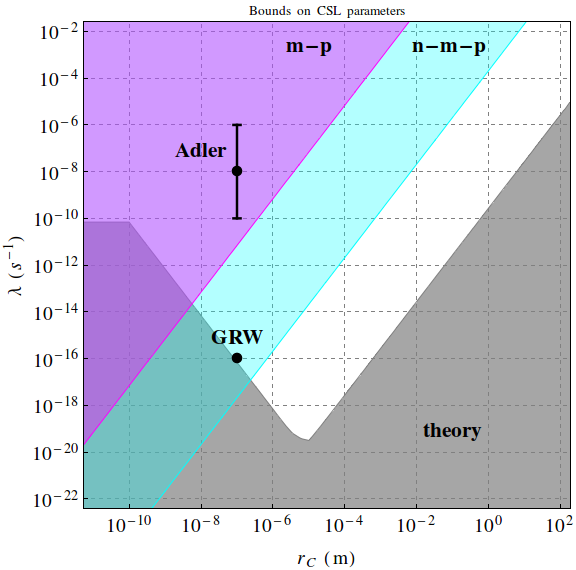}
\caption{
Mapping of the $\lambda - r_C$ Continuous Spontaneous Localization (CSL) parameters: the originally proposed theoretical values \mbox{(GRW, Adler)} are shown as black points; the region excluded by theory (theory) is represented in gray. The excluded region according to our analysis is shown in cyan for the non-mass proportional case (n-m-p) and in magenta for the mass proportional case (m-p).  
}
\label{csl}
\end{adjustwidth}
\end{figure}

\section{Conclusions and Perspectives}\label{conclusions}

We have presented an analysis of the spontaneous radiation emitted and measured by the IGEX Germanium detector, to obtain a mapping of the CSL collapse model parameters. The results shown in Figure \ref{csl} can be summarized as follows:

\begin{itemize}
\item the non-mass proportional model for a white noise scenario can be excluded by our analysis,
\item the higher value on $\lambda$ \cite{Adler} can be excluded for a white noise scenario, in both mass proportional and non-mass proportional models,
\item the measurement of the spontaneous radiation allows the obtainment of the most stringent limits on the CSL collapse model parameters, with respect to any other method, in a broad range of the parameter space (see also Ref. \cite{matteo} for comparison).
\end{itemize}

We are presently exploring the possibility of performing a new measurement that will allow an improvement of at least one order of magnitude on the collapse rate parameter $\lambda$, exploring new regions of CSL mapping.

\section*{Acknowledgments}
We acknowledge the support of the CENTRO FERMI - Museo Storico della Fisica e Centro Studi e Ricerche `Enrico Fermi' (\emph{Open Problems in Quantum Mechanics project}), the support from the EU COST Action CA 15220 is gratefully acknowledged.
Furthermore, this paper was made possible through the support of a grant from the
Foundational Questions Institute, FQXi ``Events'' as we see
them: experimental test of the
collapse models as a solution of the measurement problem) and a grant from the John Templeton
Foundation (ID 58158). The opinions expressed in this publication are those of the authors
and do not necessarily reflect the views of the John Templeton
Foundation.
Beatrix C. Hiesmayr acknowledges gratefully the support by the Autrian Science Found (FWF-P26783).
S. Donadi acknowledges the support by Trieste University and Istituto Nazionale di Fisica Nucleare (INFN).

\end{adjustwidth}
\begin{adjustwidth}{-1.5in}{0in}


\end{adjustwidth}
\end{document}